\documentclass{elsart}
\usepackage{graphicx,color}
\newlength{\parindentEnum}
\begin{document}
%*****************************************************************************
%23456789012345678901234567890123456789012345678901234567890123456789012345678
%*****************************************************************************
\begin{frontmatter}
\title{
Energy determination in the Akeno Giant Air Shower Array experiment 
}

\author[RIKEN]{M. Takeda}, 
\author[RIKEN]{N. Sakaki}, 
\author[YAMANASHIen]{K. Honda}, 
\author[KINKI]{M. Chikawa}, 
\author[ICRR]{M. Fukushima}, 
\author[ICRR]{N. Hayashida}, 
\author[SAITAMA]{N. Inoue}, 
\author[MUSASHI]{K. Kadota}, 
\author[TITECH]{F. Kakimoto}, 
\author[NISHINA]{K. Kamata}, 
\author[HIROSAKI]{S. Kawaguchi}, 
\author[OSAKA]{S. Kawakami}, 
\author[RIKEN]{Y. Kawasaki}, 
\author[YAMANASHIed]{N. Kawasumi}, 
\author[SAITAMA]{A. M. Mahrous}, 
\author[ICRR]{K. Mase}, 
\author[EHIME]{S. Mizobuchi},
\author[ICRR]{Y. Morizane}, 
\author[FUKUI]{M. Nagano}, 
\author[ICRR]{H. Ohoka}, 
\author[ICRR]{S. Osone}, 
\author[ICRR]{M. Sasaki}, 
\author[TUSOKEN]{M. Sasano}, 
\author[RIKEN]{H. M. Shimizu}, 
\author[ICRR]{K. Shinozaki}, 
\author[ICRR]{M. Teshima}, 
\author[ICRR]{R. Torii}, 
\author[YAMANASHIed]{I. Tsushima}, 
\author[RADIO]{Y. Uchihori}, 
\author[ICRR]{T. Yamamoto}, 
\author[CHIBA]{S. Yoshida}, and 
\author[EHIME]{H. Yoshii} 

\address[RIKEN]{
RIKEN (The Institute of Physical and Chemical Research), 
Saitama 351-0198, Japan
}
\address[YAMANASHIen]{
Faculty of Engineering, Yamanashi University, Kofu 400-8511, Japan
}
\address[KINKI]{
Department of Physics, Kinki University, Osaka 577-8502, Japan
}
\address[ICRR]{
Institute for Cosmic Ray Research, University of Tokyo, Chiba 277-8582, Japan
}
\address[SAITAMA]{
Department of Physics, Saitama University, Urawa 338-8570, Japan
}
\address[MUSASHI]{
Faculty of Engineering, Musashi Institute of Technology, 
Tokyo 158-8557, Japan 
}
\address[TITECH]{
Department of Physics, Tokyo Institute of Technology, Tokyo 152-8551, Japan
}
\address[NISHINA]{
Nishina Memorial Foundation, Komagome, Tokyo 113-0021, Japan
}
\address[HIROSAKI]{
Faculty of Science and Technology, Hirosaki University, 
Hirosaki 036-8561, Japan
}
\address[OSAKA]{
Department of Physics, Osaka City University, Osaka 558-8585, Japan
}
\address[YAMANASHIed]{
Faculty of Education, Yamanashi University, Kofu 400-8510, Japan
}
\address[EHIME]{
Department of Physics, Ehime University, Matsuyama 790-8577, Japan
}
\address[FUKUI]{
Department of Space Communication Engineering, Fukui University of Technology, 
Fukui 910-8505, Japan
}
\address[TUSOKEN]{
Communications Research Laboratory, Ministry of Posts and Telecommunications, 
Tokyo 184-8795, Japan
}
\address[RADIO]{
National Institute of Radiological Sciences, Chiba 263-8555, Japan
}
\address[CHIBA]{
Department of Physics, Chiba University, Chiba 263-8522, Japan
}

%*****************************************************************************
%23456789012345678901234567890123456789012345678901234567890123456789012345678
%*****************************************************************************
\pagebreak
\begin{abstract}
%% 1990Feb  1997Oct(PRL)   7+2/3
%%          2000May(LANL) 10+1/4
Using data from more than ten-years of observations 
with the Akeno Giant Air Shower Array (AGASA), 
we published a result that 
the energy spectrum of ultra-high energy cosmic rays 
extends beyond the cutoff energy predicted by Greisen \cite{gzk_g}, 
and Zatsepin and Kuzmin \cite{gzk_zk}. 
In this paper, we reevaluate 
the energy determination method used for AGASA events 
%with accumulated data 
with respect to the lateral distribution of shower particles, 
their attenuation with zenith angle, shower front structure, 
delayed particles observed far from the core and other factors.
The currently assigned energies of AGASA events 
have an accuracy of 
$\pm$25\% in event-reconstruction resolution and 
$\pm$18\% in systematic errors around 10$^{20}$eV. 
This systematic uncertainty is independent of primary energy 
above 10$^{19}$eV. 
Based on the energy spectrum from 10$^{14.5}$eV to a few times 10$^{20}$eV
determined at Akeno, 
there are surely events above 10$^{20}$eV and the energy
spectrum extends up to a few times 10$^{20}$eV without a GZK-cutoff.
\end{abstract}

\begin{keyword}
Extensive air showers \sep
ultra high energy cosmic rays \sep 
energy determination 
\PACS 
96.40.Pq \sep
95.55.Vj \sep
96.40.De \sep
95.85.Ry
\end{keyword}
\end{frontmatter}
%*****************************************************************************
%23456789012345678901234567890123456789012345678901234567890123456789012345678
%*****************************************************************************
\section{Introduction}

From ten-years of data collected by
the Akeno Giant Air Shower Array (AGASA), 
we have shown that 
the energy spectrum of primary cosmic rays extends 
up to a few times 10$^{20}$eV 
without the expected GZK cutoff \cite{takeda98a}. 
On the other hand, 
the HiRes collaboration has recently claimed that 
the GZK cutoff may be present with their exposure being 
similar to AGASA \cite{HiResICRC2001}. 
Ave et al. \cite{ave01a} have re-analyzed the Haverah Park events and 
their energies are reduced by about 30\% 
using a new energy conversion formula.
Although we have already published our statistical and systematic 
errors in the energy determination in related papers 
\cite{takeda98a,hayashida94a,yoshida94a,yoshida95a}, 
it is now quite important to reevaluate uncertainties in 
the energy determination of AGASA events 
with the accumulated data of ten years. 
The uncertainties due to shower front structure and 
delayed particles far from a shower core are also 
evaluated and described in some detail.

The AGASA array is the largest operating surface array, 
covering an area of about 100km$^{2}$ and 
consisting of 111 surface detectors of 2.2m$^{2}$ area 
\cite{teshima86a,chiba92a}. 
Each surface detector is situated with a nearest-neighbor separation 
of about 1km  and 
the detectors are sequentially connected with pairs of optical fibers. 
All detectors are controlled at detector sites with their own CPU and 
through rapid communication with a central computer. 
In the early stage of our experiment 
AGASA was divided into four sub-arrays called ``branches'' 
for topographical reasons, 
and air showers were observed independently in each branch. 
The data acquisition system of AGASA was improved 
and the four branches were unified into a single detection system 
in December 1995 \cite{ohoka97a}. 
After this improvement 
the array has operated in a quite stable manner
with a duty cycle of about 95\%, 
while the duty cycle before unification was 89\%. 

In a widely spread surface array like AGASA, 
the local density of charged particles 
at a specific distance from the shower axis 
is well established as an energy estimator \cite{hillas71a} 
since the local density of the electromagnetic component 
depends weakly on variations in interaction models, 
fluctuations in shower development and primary mass. 
In the AGASA experiment, 
we adopt the local density at 600m, $S(600)$, 
which is determined by fitting 
a lateral distribution function (LDF) of observed particle densities 
to an empirical formula \cite{yoshida94a}. 
This empirical formula is found to be valid for 
EAS with energies up to 10$^{20}$eV 
and for zenith angles smaller than 45$^{\circ}$ \cite{doi95a,sakaki99a}. 
The relation for converting $S(600)$ to primary energy 
has been evaluated so far by Monte Carlo simulations \cite{dai88a} 
up to 10$^{19}$eV and is 
\begin{equation}
	E = 2.03 \times 10^{17} \cdot S_{0}(600) \hspace{1em} \mbox{eV}
	\hspace{2em}, 
\label{eq:econv}
\end{equation}
where $S_{0}(600)$ is the $S(600)$ value per m$^{2}$ 
for a vertically incident shower. 
This conversion relation is derived from electron components 
for air showers observed 900m above sea level. 
In \S \ref{ssect:energy}, 
a new conversion constant is evaluated 
taking account of the average altitude. 
More modern simulation codes have been used in this simulation. 
%***** Don't understand the rest of this sentence ``AGASA single particle''.  

In the southeast corner of AGASA there is the Akeno 1km$^2$ array
\cite{hara79a}.  This is a densely packed array of detectors
covering an area of 1km$^2$ operated since 1979.  This array was
used to determine the energy spectrum between 10$^{14.5}$eV and
10$^{18.5}$eV.  In this experiment, the total number of
electrons, known as the shower size $N_e$, was used as an energy
estimator.  The relation between this energy spectrum and the
AGASA energy spectrum is discussed in \S \ref{sect:spectrum}.

%*****************************************************************************
\section{Densities measured by scintillation detectors}
\label{sect:density}

The AGASA array consists of plastic scintillators of 2.2m$^{2}$ area, 
and the light from these scintillators 
is viewed by a 125mm diameter Hamamatsu R1512 
photomultiplier tube (PMT) at the bottom of each enclosure box. 
The scintillators are 5cm thick (0.14 radiation lengths). 
The enclosure box and a detector hut are made of steel 
with 2mm and 0.4mm thickness, respectively. 
There is 
another type of enclosure box 
used for 17 detectors in the Akeno branch, 
in which PMT is mounted at the top of each enclosure box. 

\begin{figure}
\includegraphics[width=0.95\textwidth]{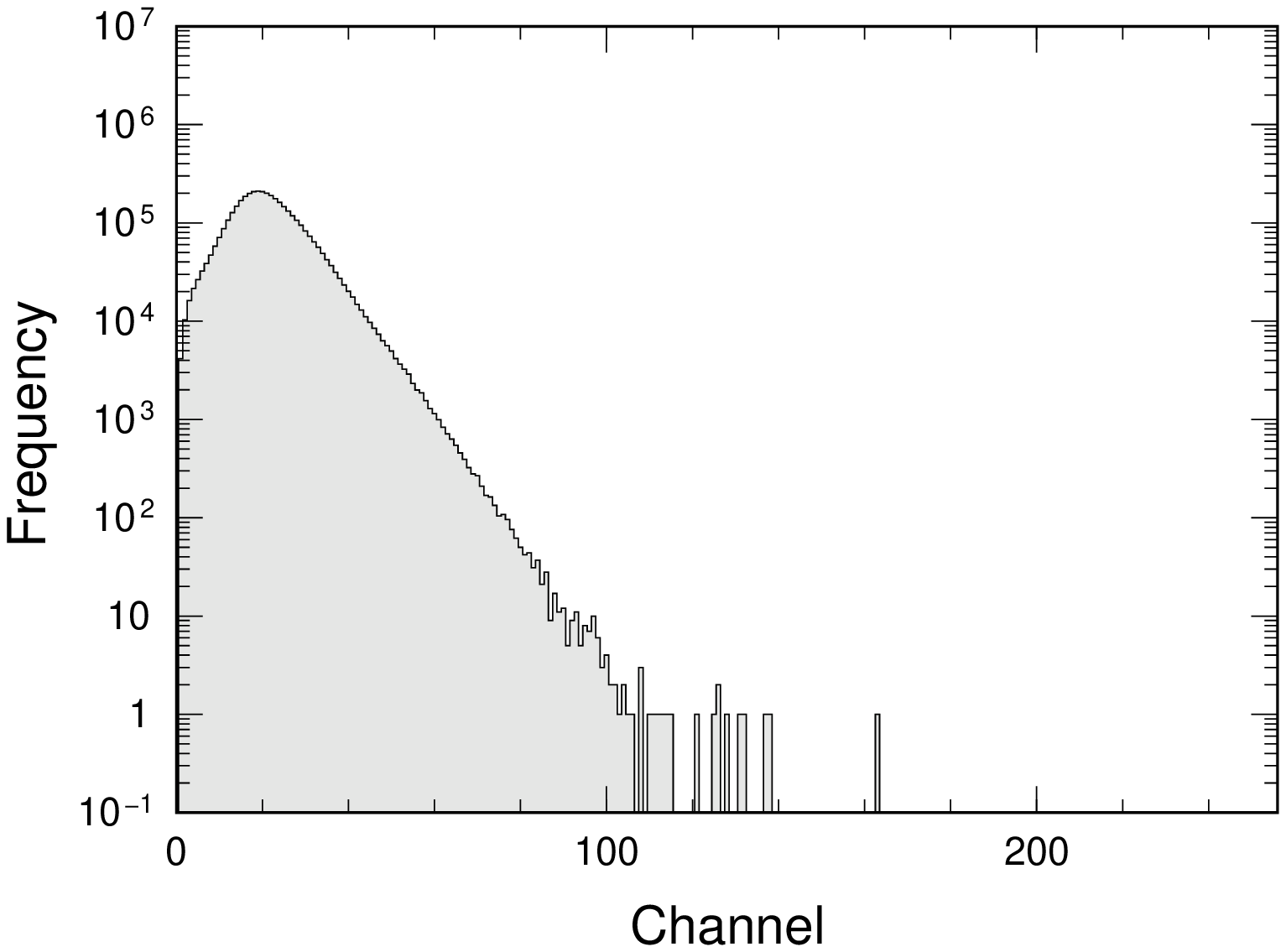}
\caption{
An example of the pulse width distribution 
of a scintillation detector (TB22) for one run (about half a day). 
One channel corresponds to 500ns. 
This distribution is used to monitor the gain and 
the decay time of the exponential pulse. 
}
\label{fig:tb22_pwd}
\end{figure}

In order to cover a dynamic range 
from 0.3 to a few times 10$^{4}$ particles 
per detector, a logarithmic amplifier is used \cite{teshima86a}. 
The number of incident particles is determined from the pulse width, 
which is obtained by presenting such a signal 
that decays exponentially with a time constant of $\tau \simeq 10\mu$s 
to a discriminator with a constant threshold level $V_{d}$. 
The relation between the number of incident particles $N$ 
and the pulse width $t_{d}$ is given by 
\begin{equation}
	V_{d} = V e^{-t_{d}/\tau} = k N e^{-t_{d}/\tau} \hspace{2em} ,
\label{eq:logamp}
\end{equation}
where $V$ is the pulse height ($V = k N$, $k$ is a constant 
depending on the gain of the amplifier). 
By defining the pulse width for $N = 1$ as $t_{1}$, 
one obtains 
\begin{equation}
	\ln N = \frac{t_{d} - t_{1}}{\tau} \hspace{2em} . 
\label{eq:log_number}
\end{equation}
Figure \ref{fig:tb22_pwd} shows a typical pulse width distribution (PWD)
for omni-directional muons, and 
its peak value is used
as $t_{1}$ 
in the experimental convenience. 

In the Akeno experiment, 
the original definition of a ``single particle'' was based on 
the average value ($PH_{ave}^{0}$) of a pulse height distribution (PHD) 
from muons traversing a scintillator vertically \cite{nagano84a}. 
This $PH_{ave}^{0}$ is accidentally coincident with 
the peak value $PH_{peak}^{\theta}$ of the PHD of omni-directional muons, 
since the PHD is not a Gaussian distribution 
but is subject to Landau fluctuations 
and coincidental incidence of two or more particles. 
If we express the pulse height corresponding to the peak value
of PWD as $PW_{peak}^{\theta}$, 
the parameter $PH_{peak}^{\theta}$ is 
related to $PW_{peak}^{\theta}$ by
\begin{equation}
	PW_{peak}^{\theta} = 
		\frac{1}{2}
		\left( PH_{peak}^{\theta} + 
			\sqrt{{PH_{peak}^{\theta}}^{2} + 4 \sigma^{2}}
		\right)
	\hspace{2em} .
\label{eq:pwph}
\end{equation}
This equation can be derived under an assumption 
that main part of PHD is expressed by a Gaussian distribution 
with a standard deviation of $\sigma$.
By converting the pulse height of this Gaussian distribution to 
the pulse width by using Equation (2), and by evaluating $t_1$ 
at the maximum of the distribution, Equation (4) is obtained. 
With $PH_{peak}^{\theta} = 1.0$ and $\sigma = 0.35$,
$PW_{peak}^{\theta} = 1.1$. 
The density measured in units of $PW_{peak}^{\theta}$, therefore, 
is 1.1 times smaller than that 
measured in units of $PH_{peak}^{\theta} (= PH_{ave}^{0})$. 
On the other hand, the density measured with a scintillator 
in units of $PH_{ave}^{0}$ 
is 1.1 times 
larger than the particle density measured with spark chambers 
between 10m and 100m from shower cores \cite{shibata65a}. 
The efficiencies and other details of the spark chamber
are described in \cite{nagano65}.
This factor 1.1 was interpreted as due to the
transition effect of electromagnetic components in scintillator
compared to the electron density 
measured in spark chamber  \cite{shibata65a}.

This means that the density in units of $PW_{peak}^{\theta}$ 
corresponds to an electron density measured by a spark chamber, 
given that the ratio of densities measured with 
scintillators and spark chambers is 1.1 . 
The number of particles in units of $PW_{peak}^{\theta}$, therefore, 
coincides with the electron density and has been applied so far 
to estimate primary energy using Equation (\ref{eq:econv}).

\begin{figure}
\includegraphics[width=\textwidth]{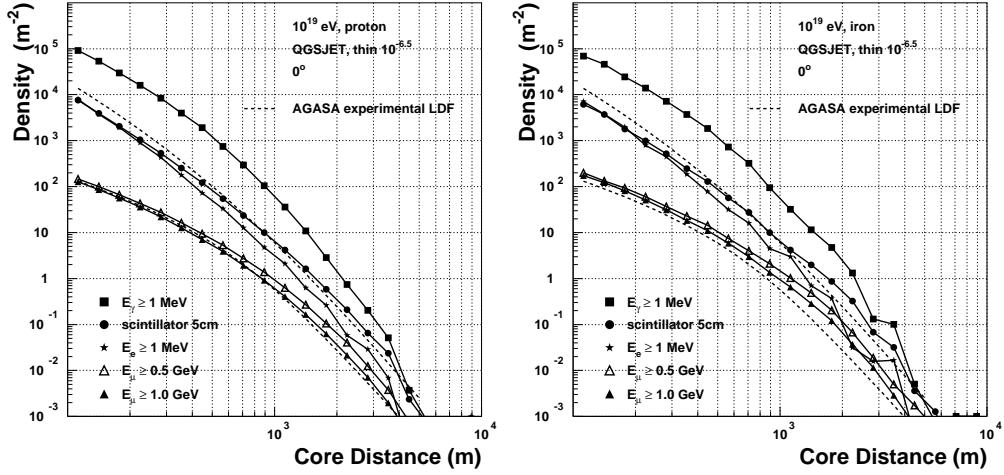}
\caption{
The lateral distribution of energy deposited in an AGASA scintillator ($\bullet$) 
in units of $PH_{ave}^{0}$ is compared with the experimental 
lateral distribution of AGASA (dashed curve). 
The density of electrons ($\geq$ 1MeV), photons ($\geq$ 1MeV), 
muons ($\geq$ 0.5GeV) and muons ($\geq$ 1GeV) are also plotted. 
(Left: proton primary; Right: iron primary)
[From Nagano et al., 1998]
}
\label{fig:detsim_lateral}
\end{figure}

To examine whether a ``single particle'' is appropriate, 
the CORSIKA program has been used to simulate densities 
measured by a scintillator of a 5cm thickness \cite{nagano00a}. 
In this simulation, a ``single particle'' corresponds to $PH_{ave}^{0}$. 
Figure \ref{fig:detsim_lateral} plots 
the lateral distribution of energy deposit in the scintillator 
in units of $PH_{ave}^{0}$ (closed circles), and 
it is compared with the experimental LDF (dashed curve). 
The simulated LDF is flatter than the experimental one. 
The simulated density reflects the number of electrons 
near the core (up to about 200m from the core), 
but becomes larger than the electron density with increasing core distance. 
Recently we have also studied the detector response with the
GEANT simulation \cite{sakaki01a,sakaki01b}. 
In this simulation, a ``single particle'' is defined as the peak value 
of $\log_{10}$(energy deposit in scintillator) for 
omni-directional muons with their energy spectrum 
to represent the experimental $PW_{peak}^{\theta}$. 
Here we take account of the real configuration of a detector, 
conversion of photons in the wall of the enclosure box and the detector hut, 
scattering of particles, decay of unstable particles (pions, kaons and etc), 
and the 4-momentum of shower particles. 
The shape of the lateral distribution is nearly consistent with 
the experimental LDF, 
though it is also a little flatter than the experimental one. 
The details will be described in \cite{sakaki01b}.

%*****************************************************************************
\section{Evaluation of uncertainties on energy estimation in AGASA}

\subsection{Detector}					%*********************
\label{ssect:detector}

\begin{figure}
\includegraphics[width=0.95\textwidth]{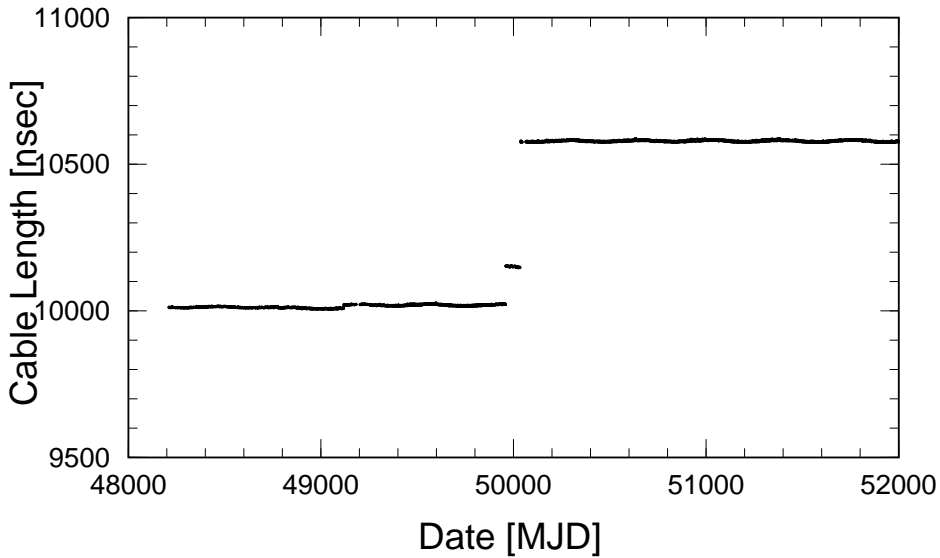}
\caption{
An example of time variation of the cable delay (in nanoseconds) 
between a detector (TB22) and the control unit at a branch center. 
The discontinuity around 50,000 MJD is 
due to a change in the detector position and 
another one is due to a system upgrade in 1995. 
}
\label{fig:tb22_cable}
\end{figure}

The detector positions were measured using a stereo camera from an airplane 
with accuracies of $\Delta X, \Delta Y =$ 0.1m and $\Delta Z =$ 0.3m. 
The cable lengths (the propagation delay times of signals) 
from the Akeno Observatory to each detector is 
regularly measured 
with accuracy of 0.1ns in each RUN (about twice a day). 
Figure \ref{fig:tb22_cable} shows 
the variation in cable length for a typical detector 
as a function of day. 
A discontinuity around 50,000 MJD is 
due to the movement of the detector position and 
another one is due to the system upgrade in 1995. 

\begin{figure}
\includegraphics[width=\textwidth]{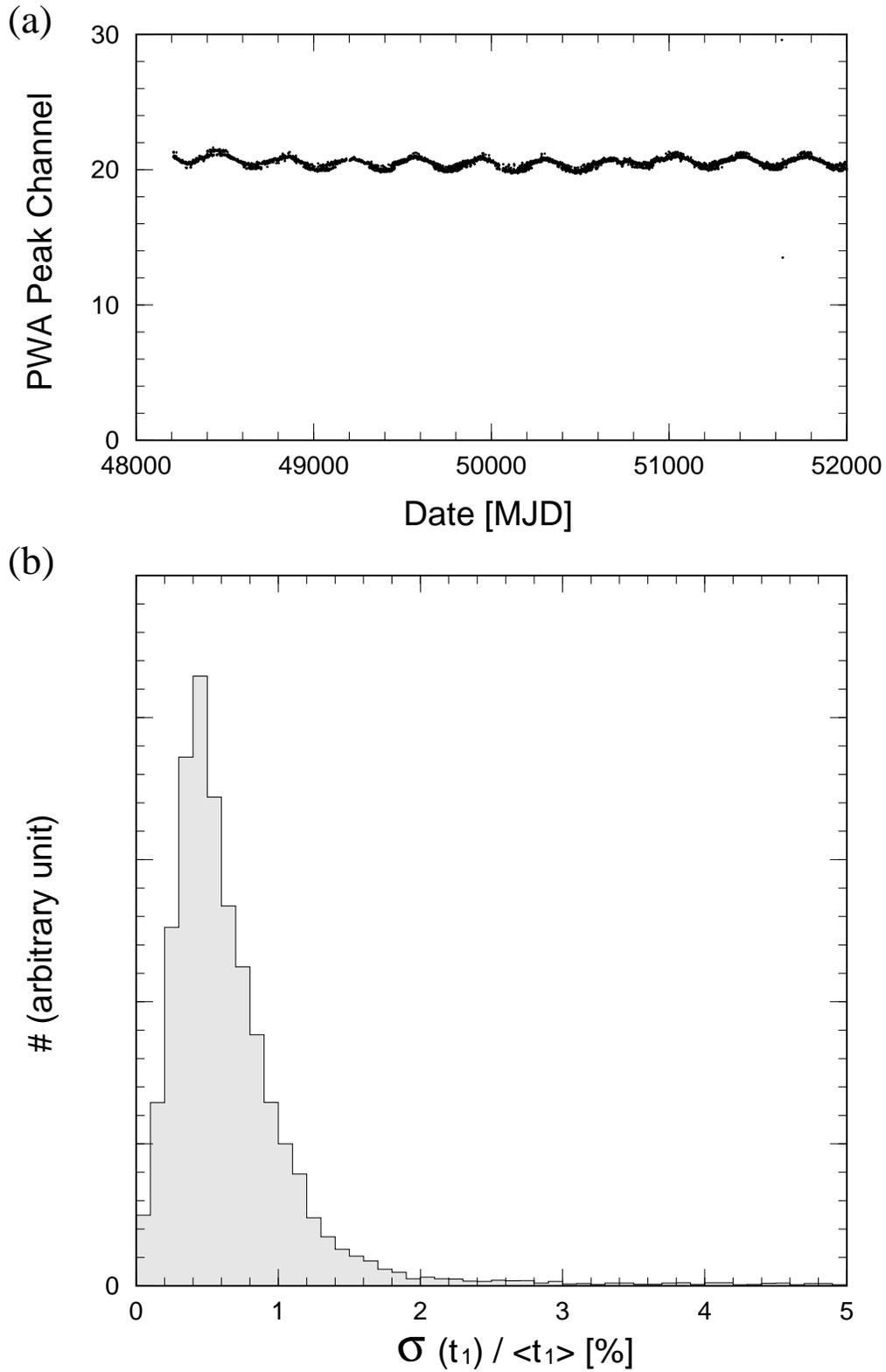}
\caption{
(a) An example of the time variation in the PWA peak channel ($t_{1}$) 
    of a scintillation detector (TB22). 
(b) Distribution of $\sigma(t_{1}) / \langle t_{1} \rangle$ 
    for all detectors, determined with the data from one month. 
}
\label{fig:tb22_peak}
\end{figure}

In Equation (\ref{eq:log_number}), there are two parameters 
which should be determined. 
The first one is the ``single particle'' $t_{1}$ ($= PW_{peak}^{\theta} $). 
In the AGASA experiment, 
pulse widths of all incident particles are recorded and 
their PWD is stored in the memory as shown in Figure \ref{fig:tb22_pwd}, 
and then $t_{1}$ is determined in every RUN. 
Figure \ref{fig:tb22_peak}(a) shows the time variation of $t_{1}$ 
for a typical detector over 11-years of operation. 
There is a clear seasonal variation with a $\pm$3\% fluctuation, 
but this variation has been calibrated 
in the air shower analysis using monthly data. 
The variance $\sigma^{2}(t_{1})$ within each month of data 
is determined and $\sigma(t_{1})$ 
is shown for all detectors in Figure \ref{fig:tb22_peak}(b),  
with $\sigma(t_{1}) / \langle t_{1} \rangle \leq 0.7\%$ for a 68\% C.L. 

\begin{figure}
\includegraphics[width=\textwidth]{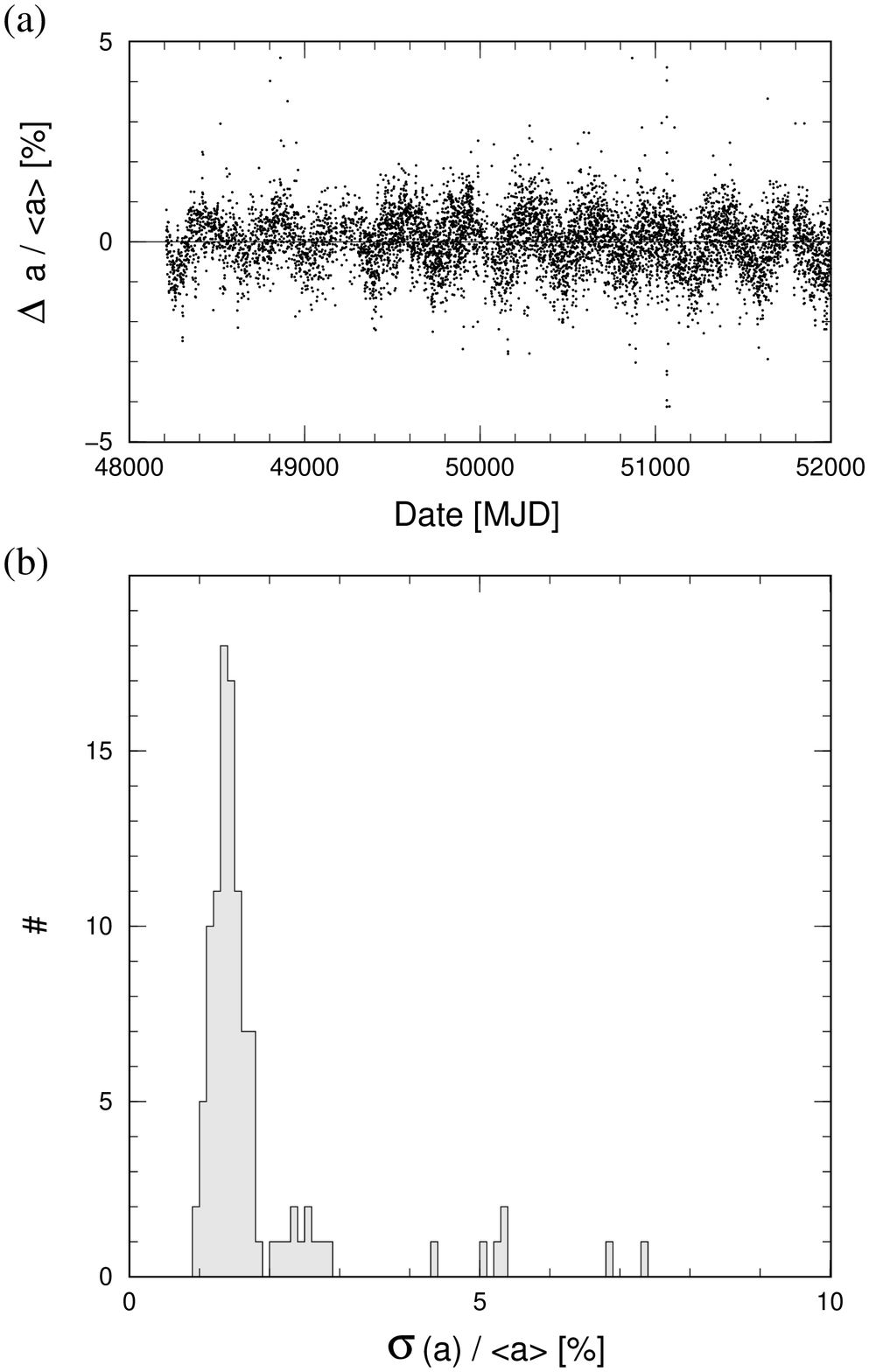}
\caption{
(a) Time variation of $\Delta a / a$ for a detector (TB22). 
(b) Distribution of $\sigma(a) / \langle a \rangle$ for all detectors, 
    determined with data from one month. 
}
\label{fig:tb22_slope}
\end{figure}

The second important parameter is the decay constant $\tau$. 
Although we have directly measured the $\tau$ values 
with a LED 
several times during AGASA's operation, 
they are not enough to estimate the time variation of $\Delta \tau / \tau $. 
We estimate this variation using the observed PWDs. 
Assuming the density (particle number) spectrum of 
incident particles in a detector is 
$ I \propto N^{- \gamma} $ ,
one obtains 
\begin{equation}
	\Delta \ln I 
		=  - \gamma \frac{\Delta N}{N}
		= - \gamma \frac{\kappa \Delta x}{\tau} 
	\hspace{3em} (t_{d} = \kappa \, x)
	\hspace{2em} .
\end{equation}
The ratio $a \equiv \Delta \ln I / \Delta x$ is the slope of the PWD, 
so that the ratio $\Delta \tau / \tau $ is expressed by 
\begin{equation}
	\frac{\Delta \tau}{\tau} = - \frac{\Delta a}{a}
	\hspace{2em} . 
\end{equation}
Figure \ref{fig:tb22_slope}(a) shows 
the time variation in $\Delta a / a$ for a typical detector. 
The variance $\sigma^{2}(\frac{\Delta a}{a})$ is determined
throughout the observation time (11 years) for each detector.
For all 111 detectors, $\sigma(\frac{\Delta a}{a})$ is plotted in 
Figure \ref{fig:tb22_slope}(b), 
and 
$\sigma(\frac{\Delta a}{a}) / \langle \frac{\Delta a}{a} \rangle \leq 1.6\%$ 
at a 68\% C.L. 
This fluctuation causes an uncertainty in density estimation 
of  4\%, 7\%, 11\% and 15\% for 10, 10$^{2}$, 10$^{3}$ and 10$^{4}$ 
particles per detector, respectively. 
It should be noted that 
$\sigma(\frac{\Delta a}{a})$ includes not only the change of $\tau$ 
but also that of $\gamma$ caused by varying atmospheric conditions 
(temperature and pressure). 
The real variation in $\Delta \tau / \tau$ is, therefore, 
smaller than that for $\Delta a / a$ discussed above.

\subsection{Air Shower Phenomenology}			%*********************
\label{ssect:phenomenology}

\begin{figure}
\includegraphics[width=\textwidth]{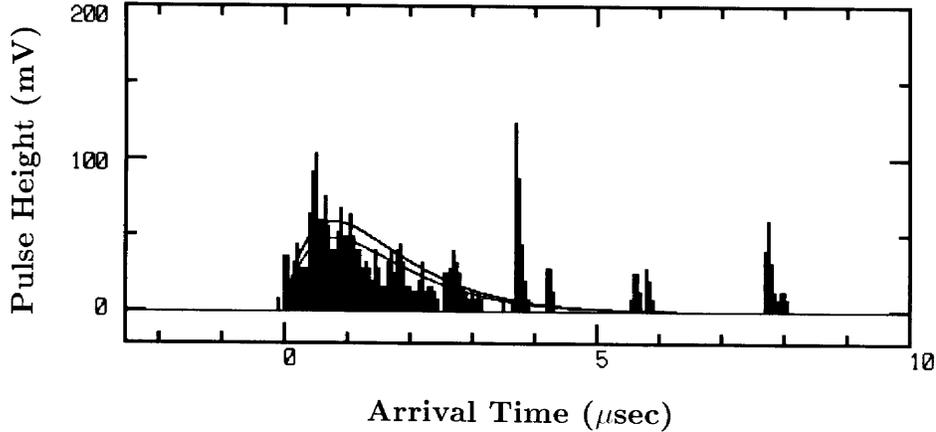}
\caption{
Arrival time distribution of charged particles in a 30m$^{2}$ detector 
measured by a wave form recorder at 1,920m from the core 
for a 2 $\times$ 10$^{20}$eV event. 
Solid curves correspond to $t_{0} =$ 800ns. 
The areas are normalized to the number of particles 
within 2.5$\mu$s (87 particles) and 3.5$\mu$s (115 particles), respectively. 
[From Hayashida et al., 1994]
}
\label{fig:yamanashi30m2}
\end{figure}

\begin{enumerate}
\setlength{\parindent}{\parindentEnum}
\item	Lateral Distribution Function:

	In general, the number of detectors with measurable
	particle density is not enough to treat both the core
	location and the slope parameter in lateral distribution
	as free parameters for most of the observed showers.
	In order to avoid systematic errors in determination 
	of lateral distribution, we have used only showers well
	within the boundary of the array and introduced a 
	new parameter which represents the degree of goodness
	in locating the core position which is estimated by the
	maximum likelihood method. The details is described 
	in \cite{yoshida94a}.  

	The empirical LDF thus determined \cite{yoshida94a} is expressed by 
	\begin{equation}
		\rho(r) \propto \left( \frac{r}{R_{M}} \right)^{- 1.2}
			\left( 1 + \frac{r}{R_{M}} \right)^{- (\eta - 1.2)}
			\left\{ 1 + \left( \frac{r}{1000} \right)^{2}
			\right\}^{-0.6}		\hspace{2em} ,
	\label{eq:ldf}
	\end{equation}
	where $r$ is the distance from the shower axis in meters. 
	The Moliere unit $R_{M}$ is 91.6m at the Akeno level. 
	The parameter $\eta$ 
	is a function of zenith angle $\theta$ expressed by 
	\begin{equation}
		\eta = (3.97 \pm 0.13) - (1.79 \pm 0.62) \, (\sec \theta - 1)
		\hspace{2em} .
	\label{eq:eta}
	\end{equation}
	The uncertainty in the energy determination of showers 
	due to the limited accuracy in determination of $\eta$ 
	was discussed and estimated to be $\pm$10\% 
	by Yoshida et al. \cite{yoshida94a}. 

	With observed showers, 
	we have confirmed in \cite{sakaki99a} that 
	the empirical formula of Equations (\ref{eq:ldf}) 
	can be applied 
	to showers with energies up to 10$^{19.8}$eV 
	and with core distances up to 3km 
	by improving the errors of Equation (\ref{eq:eta}) by 
	\begin{equation}
		\eta = (3.84 \pm 0.11) - (2.15 \pm 0.56) \, (\sec \theta - 1)
		\hspace{2em} .
	\label{eq:etarev}
	\end{equation}
	We may extend the present lateral distribution up to
	the highest energy observed, since any energy dependence
	of $\eta$ has not been observed.
	In the same manner as Yoshida et al. \cite{yoshida94a}, 
	the systematic effect on $S(600)$ estimation 
	due to uncertainties in Equation (\ref{eq:etarev}) is 
	evaluated to be $\pm$7\% 
	for air showers with zenith angles smaller than 45$^{\circ}$. 

\item	Atmospheric Attenuation: 

	Since an inclined air shower traverses the atmosphere deeper 
	than a vertical shower, a shower density
	$S_{\theta}(600)$ observed at zenith angle $\theta$ must be 
	transformed into $S_{0}(600)$ corresponding to a vertical shower. 
	The attenuation of $S_{\theta}(600)$ 
	is formulated as follows: 
	\begin{equation}
		S_{\theta}(600) = S_{0}(600) \, \exp \left[
		- \frac{X_{0}}{\Lambda_{1}} (\sec \theta - 1)
		- \frac{X_{0}}{\Lambda_{2}} (\sec \theta - 1)^2
		\right]
		\hspace{2em} ,
	\label{eq:attenuation}
	\end{equation}
	where $X_{0} =$ 920g/cm$^{2}$, 
	$\Lambda_{1} = 500$g/cm$^{2}$ and 
	$\Lambda_{2} = 594^{+268}_{-120}$g/cm$^{2}$ 
	for $\theta \leq 45^{\circ}$ \cite{yoshida94a}. 
	The uncertainty in $S_(600)$ determination 
	due to the uncertainty in the attenuation curve 
	of $S(600)$ was also discussed there. 

	The attenuation curve of $S(600)$ 
	is now under reevaluation with
	the accumulated data up to zenith angles $\theta \leq 60^{\circ}$ 
	and using a modern simulation code. 
	Equation (\ref{eq:attenuation}) is valid 
	for observed events with $\theta \leq 45^{\circ}$, 
	and AIRES simulation agrees well 
	with the experimental formula and data \cite{sakaki99a}. 
	Another important point of this study is that 
	$\Lambda$s are independent of energy. 
	The uncertainty in $S_{0}(600)$ due to this transformation 
	is estimated to be $\pm$5\%;
	this value is also reduced from Yoshida et al. \cite{yoshida94a}
	because of the increased amount of observed showers. 

	In the analysis 
	we have assumed the symmetric distribution of particles 
	around the shower axis 
	by neglecting the difference of attenuation to each detector. 
	The experimental LDF is determined by the average of 
	these densities, and the fluctuation around the average is 
	determined with these densities.
	In analysis procedure described in \S \ref{ssect:analysis}, 
	this fluctuation is included in $\sigma$, 
	which is used as a weight of each detector.

\item	Accidental Coincidence: 

	Because we use the log-amplifier described above,
	the density could be overestimated 
	if an accidental signal hits on the tail of the exponential pulse 
	above the threshold level of the discriminator.
	The counting rate of the scintillation detector (area 2.2m$^2$) 
	is about 500Hz for signals exceeding 
	the threshold of 0.3 particles per detector. 
	The accidental coincidence of a background particle hitting
	a detector within the pulse of a single particle (10$\mu$s width) 
	occurs with a chance probability of 
	$ 10 \times 10^{-6} \times 500 = 5 \times 10^{-3} $. 
	In the same way, 
	one obtains 
	1.65 $\times$ 10$^{-2}$ for the probability 
	during a 10 particle pulse (33$\mu$s width) 
	and 5.6 $\times$ 10$^{-2}$ for a 100 particle pulse (56$\mu$s width). 
	This means that one of 200, 61, or 18 detectors in each case 
	may record larger values than the real density. 
	Frequency of accidental pulses is much less than
	that of delayed particles described in \S \ref{ssect:phenomenology}(e) 
	and their pulse heights are smaller than 1 particle.  
	With the present analysis method described in \S \ref{ssect:analysis}, 
	a detector which deviates more than 3$\sigma$ from the average LDF 
	is excluded and hence 
	this effect is negligible. 

\item	Shower Front Structure: 

	Given our use of the log-amplifier, 
	the density is properly estimated 
	so long as the thickness of a shower front 
	is less than a few 100ns. 
	However, if the thickness is larger than this time width, 
	we should take this effect into account 
 	to estimate the incident particle density appropriately. 
 	Figure \ref{fig:yamanashi30m2} is 
	an example of the arrival time distribution 
	observed with a 30m$^{2}$ scintillator \cite{hayashida94a}, 
	operated by the Yamanashi university group 
	and triggered by AGASA. 
	The core distance is 1,920m and 
	the primary energy is 2 $\times$ 10$^{20}$eV. 
	Not only is the particle arrival time distribution broad,
	but 5 particles are delayed more than 3$\mu$s in this 30m$^{2}$
	detector.

	The arrival time distribution of shower particles has been 
	measured with a scintillation detector of 
	12m$^{2}$ area together with the 30m$^{2}$ detector. 
	Signal sequences of arriving particles 	are recorded 
	in time bins of 50ns for the 30m$^{2}$ detector 
	and 20ns for the 12m$^{2}$ detector 
	in coincidence with the AGASA trigger. 
	The details of these experiments are described 
	in Honda et al. \cite{khonda87a,khonda97a}. 
	From these experiments, 
	the average shape of the arrival time distribution
	is expressed by 
	\begin{equation}
		f(t, r) = \frac{t}{t_{0}(r)} \exp 
			\left( \frac{-t}{t_{0}(r)} \right)
	\hspace{2em} , 
	\label{eq:showerfront}
	\end{equation}
	where the scaling parameter $t_{0}$ is 168ns, 212ns, and 311ns 
	at $r =$ 534m, 750m, and 1,050m, respectively. 
	These are shown by solid lines in Figure \ref{fig:arrtime}.
	Beyond 1,050m 
	the events are too few to determine the average $t_{0}$. 
	However, 
	if we extrapolate the relation assuming $t_{0} \propto r$, 
	we find $t_{0} =$ 440ns at $r =$ 1,500m and 570ns at 2,000m.
	If we extrapolate it assuming $\log t_{0} \propto r$, 
	$t_{0} =$ 490ns at $r =$ 1,500m and 850ns at 2,000m, respectively.
	The $t_{0}$ distribution of each event in 
	these distant ranges seems to be nearer to, 
	but shorter than the values extrapolated with $\log t_{0} \propto r$. 
	We use, therefore, 
	the latter values in the present evaluation as upper bound 
	up to 2,000m and for events with energies up to 10$^{20}$eV.
	These distributions are also drawn with
	dotted lines in Figure \ref{fig:arrtime}. 
	The solid curves in Figure \ref{fig:yamanashi30m2} correspond to 
	the time distribution with $t_{0} =$ 800ns, 
	which support the extrapolation with $\log t_{0} \propto r$. 

	Using $t_{0}$ and the number of incident particles as parameters,
	we have derived the ratio (the overestimation factor) 
	of the estimated density 
	due to the broadening of the shower front structure to
	the density with  $t_{0} = 0$ as a function
	of core distance and primary energy.
	The results are drawn in Figure \ref{fig:eff_thickness_ldf}.
	The factor is nearly independent of primary energy up to a few 1,000m. 
	The factor increases rapidly with core distance	above 1,500m, 
	but it decreases suddenly again at those core distances 
	where the observed number of shower particles is near unity. 
	From this figure, the overestimation factor for the density at 600m 
	is $+$3.5\% and 
	that around 1km is $+$6\% for 10$^{20}$eV showers. 
	With our analysis method described in \S \ref{ssect:analysis}, 
	the present $S(600)$ may be overestimated by about 5\% 
	due to the broadening of the shower front structure 
	with its fluctuation about $\pm$5\%.

\item	Delayed Particles: 
\label{it:delay} % (e)

	As shown in Figure \ref{fig:yamanashi30m2}, 
	there are particles at large core distances
	which are delayed by more than a few micro seconds 
	with respect to normal shower particles.
	It was shown in the prototype AGASA experiment 
	that pulses delayed by more than 4$\mu$s are 
	most likely to be low energy neutrons with energy 30--40MeV 
	and the fraction of these pulses to the total shower particles 
	is a few \% between 1km and 3km \cite{teshima86b}. 
	Based on this result, we have so far assumed that 
	the effect of delayed particles on the $S(600)$ determination 
	is within the errors due to other effects. 

	In the following, we evaluate the effect of delayed particles 
	on the $S(600)$ determination with accumulated data from ten years
	of operation. 
	If a delayed particle, whose energy loss in a scintillator 
	corresponds to $N_{D}$ particles, hits a detector 
	with time delay $t_{D}$ with respect to $N_{i}$ incident particles at $t = 0$, 
	the pulse height $V(t)$ is expressed by 
	\begin{eqnarray}
		V(t) & = & N_{i} \exp \left(- \frac{t}{\tau} \right)
			+ N_{D} \exp \left(- \frac{t-t_{D}}{\tau} \right)
			= OF \cdot N_{i} \exp \left(- \frac{t}{\tau} \right)
	\label{eq:of_delay}
		\\
	& &	OF = 1 + \frac{N_{D}}{N_{i}} \exp 
			\left( \frac{t_{D}}{\tau} \right)
		\nonumber
	\hspace{2em} , 
	\end{eqnarray}
	where $N_{i}$ and $N_{D}$ are in units of a ``single particle'', 
	and $OF$ is the overestimation factor due to delayed particles. 
	The $OF$ value depends on $N_{D}/N_{i}$ and $t_{D}$. 
	It is, therefore, quite important to evaluate the density
	of delayed particles and 
	their energy loss in a 5cm scintillator experimentally. 
	These values have been measured with the scintillation detectors 
	of 30m$^{2}$ and 12m$^{2}$ area 
	described in the previous section.

	Figure \ref{fig:NdvsE} shows 
	the ratio of delayed particles (delay time $t_{D} \geq 3\mu$s 
	and pulse height $N_{D} \geq 1.0$ particles) 
	to all shower particles measured 
	as a function of core distance for three energy ranges: 
	$\log$(Energy [eV]) $=$ 18.5--19.0 (open circles), 
	19.0--19.4 (open squares) and above 19.4 (closed squares). 
	In the same figure, 
	the previous Akeno result by Teshima et al. \cite{teshima86b}
	and the result by Linsley \cite{linsley} are also plotted 
	by small open and closed circles, respectively. 
	In these measurements, 
	ratios of delayed particles with delay time $t_{D} \geq 4\mu$s and
	pulse height $N_{D} \geq 3.0$ particles to all particles
	are plotted for showers of energies around 10$^{18}$eV.
	When we take account of 
	the different selection conditions 
	it may be concluded that the ratio of the number of delayed particles
	to all shower particles depends on core distance, and 
	is almost independent of primary energy 
	from 10$^{18}$eV to 10$^{20}$eV.

	Figure \ref{fig:delay_nd_ni} shows 
	the ratio $N_{D} / N_{i}$ as a function of core distance 
	observed by the 30m$^2$ or 12m$^2$ detectors. 
	Here, $N_{D}$ represents the energy loss in scintillators of 
	the delayed particles 
	(delay time $t_{D} \geq 3\mu$s and 
	pulse height $N_{D} \geq 1.0$ particles) 
	measured in units of $PH_{ave}^{0}$, 
	and $N_{i}$ is sum of all shower particles 
	also in units of $PH_{ave}^{0}$ 
	for showers in the same distance and energy bin. 
	Open circles represent the ratio for showers of 
	energies between 10$^{18.5}$eV and 10$^{19.0}$eV,
	and closed squares for energies above 10$^{19.0}$eV.

	In Figure \ref{fig:dis_delay_nd}, 
	the delay time 
	and $N_{D}$ of a delayed particle 
	(units of $PH_{ave}^{0}$) 
	are plotted as a function of core distance.  
	Since there is no appreciable difference
	for different primary energies, 
	delayed particles for all energy ranges 
	are put together in this analysis. 
	Details of the shower front structure and delayed particles 
	will be described elsewhere \cite{khonda02a}.

	Using the $N_{D}/N_{i}$ values in Figure \ref{fig:delay_nd_ni} 
	and the delay times in Figure \ref{fig:dis_delay_nd}, 
	the $OF$ values are estimated from Equation (\ref{eq:of_delay}) 
	and plotted in Figure \ref{fig:dis_of} as a function of core distance. 
	These $OF$ values are independent of primary energy. 
	This is understood as follows. 
	For 10$^{19}$eV showers, 
	the density at a core distance of 1km is 6/m$^{2}$
	(for AGASA detector $N_{i} =$ 13.2/2.2m$^{2}$), 
	so that the density overestimation $OF$ due to delayed particles
	with $N_{D} = 3$ is expected to be 1.31 for $t_{D} = 3\mu$s
	and 1.37 for $t_{D} = 5\mu$s.
	However, the density of the delayed particles is so small that 
	only one in 10 detectors around 1km 
	will be hit by delayed particles. 
	Since 
	the average density is determined by several detectors, 
	the $S(600)$ overestimation is limited to 4\%.
	On the other hand, 
	for 10$^{20}$eV showers 
	all detectors around 1km from the air shower core
	are likely to be hit by delayed particles.
	However, the density overestimation ($OF$) of each detector is 
	1.04 
	because the density of shower particles around 1km is
	large ($N_{i} = 132$) at 10$^{20}$eV.

	From our analysis procedure described in \S \ref{ssect:analysis},
	$S(600)$ may be overestimated due to delayed particles
	by about $+$5\% $\pm$5\%, 
	independent of primary energy. 
	It should be noted that 
	the AGASA LDF is consistent with 
	that from electromagnetic components and muons 
	simulated as described in \S \ref{ssect:detector}. 
	If we include the simulated results
	on low energy neutrons (delayed particles) 
	using AIRES code, the LDF becomes much 
	flatter than the observed LDF beyond 1km from the core. 
	A possible flattening of LDF due to delayed particles 
	is not observed experimentally 
	up to 3km from the core and up to 10$^{20}$eV.

\end{enumerate}

\subsection{Energy Estimator}				%*********************
\label{ssect:energy}

The particle density $S_{0}(600)$ in Equation (\ref{eq:econv}) 
is evaluated as the electron density and 
the AGASA density in units of $PW^{\theta}_{peak}$
corresponds to the electron density since the ratio of 
densities measured with scintillators and a spark chamber is 1.1
as described in \S \ref{sect:density}. 
Since this ratio is not measured beyond 100m, 
it is necessary to evaluate the conversion factor 
from $S(600)$ measured in units of the AGASA ``single particle''
to primary energy. 

The new conversion formula obtained is described 
in Sakaki et al.\cite{sakaki01b}
and listed in  Table \ref{tbl:econv}. 
In this simulation a ``single particle'' is defined as 
 $PW^{\theta}_{peak}$ in accordance with the experiment.
The energy conversion formula of Equation (\ref{eq:econv}) 
was estimated for the 900m altitude of the Akeno Observatory. 
Since the Akeno Observatory is located on a mountain side, 
core positions of most events are lower than this altitude. 
At the average 667m height of the AGASA detectors, 
the atmospheric depth is 27g/cm$^{2}$ larger than 
that at the Akeno Observatory. 
In the new simulation, this altitude is applied. 

If we evaluate the difference in a  factor $a$ in Table \ref{tbl:econv} 
due to the difference of average altitudes 900m and 667m, 
it leads to a 7\% increase at $S(600)=1$.  
That is, Equation (\ref{eq:econv}) evaluated at 900m is revised to 
\begin{equation}
	E = 2.17 \times 10^{17} \cdot S_{0}(600)^{1.0}
	\hspace{1em} \mbox{eV}
	\hspace{2em} ,
\label{eq:econv_dai_2}
\end{equation}
at 667m and the result agrees with the factor $a$ 
calculated using the QGSJET interaction model and a proton primary 
by Sakaki et al. \cite{sakaki01a} in Table \ref{tbl:econv}. 
 
In order to see 
the differences due to simulation codes and
hadronic interaction models, the simulation by
Nagano et al. \cite{nagano00a} using CORSIKA is also listed.  
In this simulation, the density in units of $PH^{0}_{peak}$
is used and the average altitude is 900m. 
Taking account of 
10\% overestimation of a ``single particle'' ($PH^{0}_{peak}$) and 
the 7\% underestimation due to differences in altitude, 
we may directly compare these results with Sakaki et al. \cite{sakaki01a}. 
In each simulation, 
the differences are within 10\% 
between QGSJET and SIBYLL hadronic interaction models and 
are within 10\% between proton and iron primaries. 
The difference due to the simulation code itself is within 5\%.

It is, therefore, reasonable to use 
a revised energy conversion formula
by taking the average of these simulation results at 667m 
for proton and iron primaries with AIRES (QGSJET, SIBYLL),
CORSIKA (QGSJET, SIBYLL) and COSMOS (QCDJET) yielding 
\begin{equation}
	E = 2.21 \times 10^{17} \cdot S_{0}(600)^{1.03}
	\hspace{1em} \mbox{eV}
	\hspace{2em} .
\label{eq:econv_sakaki}
\end{equation}
That is, the AGASA energies so far published must be shifted 
by $+$8.9\% at $2\times10^{17}$eV,
$+$12.2\% at $10^{19}$eV and +13.1\% at $10^{20}$eV.
The systematics due to the simulation codes, interaction
models, and mass composition may be within 10\%. 
The intrinsic $S(600)$ fluctuation in shower development 
is less than 6\% 
under each combination of primary mass and interaction model 
with the AIRES simulation \cite{sakaki01b}. 
This small difference among interaction models and compositions 
is an advantage of measuring $S(600)$ using scintillators, 
in which observed particles are dominated by electromagnetic components 
with a small contribution of muons. 

From the above discussion,
Equation (\ref{eq:econv}) used so far by the AGASA group gives 
the lowest limit in the conversion from $S(600)$ to primary energy. 
It may be more reasonable to increase the energies  $+$10\% $\pm$12\%.
A detailed study of this topic will be found 
in \cite{sakaki01b}.

\subsection{Analysis}				%*********************
\label{ssect:analysis}

Our analysis procedure for an air shower event is based on 
an iterative process to find the arrival direction of a primary cosmic ray 
and to search for the core location and the local density $S(600)$. 
To start, we assume an initial core location at the center of gravity 
of the density distribution of an observed event. 
Next, the arrival direction (zenith angle $\theta$, azimuth angle $\phi$) 
is determined by minimizing the $\chi^{2}$ function: 
\begin{equation}
	\chi^{2} = \frac{1}{n - 3} \sum_{i=1}^{n}
		\left[ \left\{ T_{i} 
		  - T_{f}({\bf r}_{i}, \theta, \phi) 
		  - T_{d}(\rho_{i}, R_{i}) - T_{0} \right\}^{2} 
		/ {T_{s}(\rho_{i}, R_{i})}^{2} \right]
	\hspace{1.5em} ,
\label{eq:chi2_dir}
\end{equation}
where 
$T_{i}$ is the observed time of the first particle 
incident on $i$-th detector 
(the detector location is expressed by ${\bf r}_{i}$ 
measured from the core position and 
$R_{i}$ is the distance from the shower axis, and 
$\rho_{i}$ denotes the observed density). 
Here, 
$T_{f}$ is the propagation time of the tangential plane of the shower front, 
$T_{d}$ is the average time delay of the shower particles 
from this tangential plane, 
$T_{s}$ is the average deviation of shower particles, 
and $T_{0}$ denotes the time when the core hits the ground. 
The parameters -- $T_{d,s}(\rho_{i}, R_{i})$ -- of shower front structure 
are obtained experimentally \cite{hara83a}. 
At this step, those detectors that make $\chi^{2}$ large 
are excluded in the calculation as signals 
with accidental muons. 
This exclusion is continued until $\chi^{2} \leq 5.0$. 
Usually, the number of excluded detectors is one or a few. 
In the next step, 
we search for the core location and the shower size, 
which corresponds to the normalization factor in Equation (\ref{eq:ldf}), 
to maximize the likelihood function: 
\begin{equation}
	\mbox{\L} = \prod_{i=1}^{n} \left( \frac{1}{\sigma_{i} \sqrt{2 \pi}} 
		\right)
		\cdot \exp \left[ - \frac{1}{2} \sum_{i=1}^{n} 
		\left( \frac{\rho_{i} - \rho(R_{i})}{\sigma_{i}} \right)^{2}
		\right]
	\hspace{2em} , 
\label{eq:ml_size}
\end{equation}
where $\rho_{i}$ is the electron density observed by $i$-th detector 
and $\rho(R_{i})$ is the particle density estimated from the LDF. 
The fluctuation of electron density $\sigma_{i}$ 
takes account of
fluctuations in the longitudinal development 
and the detector response. 
This fluctuation was experimentally expressed 
by Teshima et al. \cite{teshima86b}. 
At this step, we again exclude a detector 
if its observed density deviates by more than 3$\sigma$, and
we assume these signals are possibly overestimated by an
accidental coincidence or delayed particles. 
Finally, 
we estimate the local density $S_{\theta}(600)$ and 
convert it to the primary energy. 

\begin{figure}
\includegraphics[width=\textwidth]{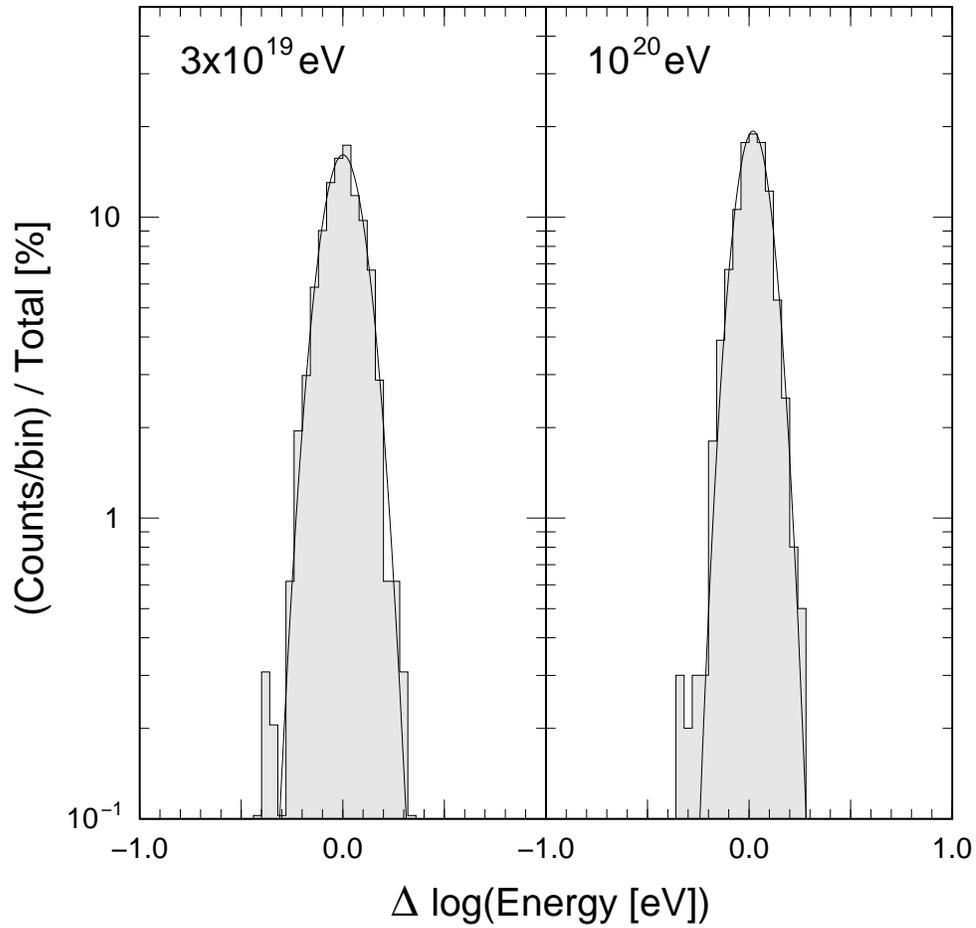}
\caption{
Accuracy of energy determination 
for 3 $\times$ 10$^{19}$eV and 10$^{20}$eV showers 
with zenith angles less than 45$^{\circ}$. 
The solid curve indicates the Gaussian distribution fitted 
to the histogram. 
}
\label{fig:resolution}
\end{figure}

Figure \ref{fig:resolution} shows 
the distributions of energies evaluated using
the above analysis method 
for a large number of artificial proton air shower events simulated  
with energies of 3 $\times$ 10$^{19}$eV and 10$^{20}$eV
at zenith angles less than 45$^{\circ}$. 
For artificial events above 10$^{19}$eV and 4 $\times$ 10$^{19}$eV, 
68\% have accuracy in arrival direction determination better 
than 2.8$^{\circ}$ and 1.8$^{\circ}$, respectively. 
These artificial events were simulated over a larger area than the AGASA area 
with directions sampled from an isotropic distribution. 
In this air shower simulation, 
the fluctuation on the longitudinal development of air showers, 
the resolution of the scintillation detectors, 
and the statistical fluctuation of observed shower particles 
at each surface detector were taken into account. 
The primary energy is determined with an accuracy of 
about $\pm$30\% at 3 $\times$ 10$^{19}$eV and $\pm$25\% at 10$^{20}$eV, 
and the fraction of events with 50\%-or-more overestimation 
in energy is only 2.4\%. 

Although only events whose cores are located within the array area are used 
in our papers, 
some events with real cores located near but outside the array boundary 
are reconstructed as ``inside'' events. 
The assigned energies of such events are smaller than their real values 
since the core distance of detectors become nearer than the true distances.
On the other hand, such events that 
are assigned "outside" the array boundary 
in the analysis procedure against their input core locations 
inside the array are excluded in our selection. 
The effects from these mislocation of cores are 
taken into account in the distributions in Figure \ref{fig:resolution} 
and the exposure in Figure \ref{fig:spec01073_fc}.

\section{AGASA energy spectrum and the relation to that 
in lower energy determined at Akeno}			%*********************
\label{sect:spectrum}

In order to derive the energy spectrum of primary cosmic rays, 
the observation time and the aperture for the selected events 
must be evaluated as a function of the primary energy. 
The aperture is determined 
by analyzing the artificial showers simulated over an area 
larger than the AGASA area described above. 

\begin{figure}
\includegraphics[width=\textwidth]{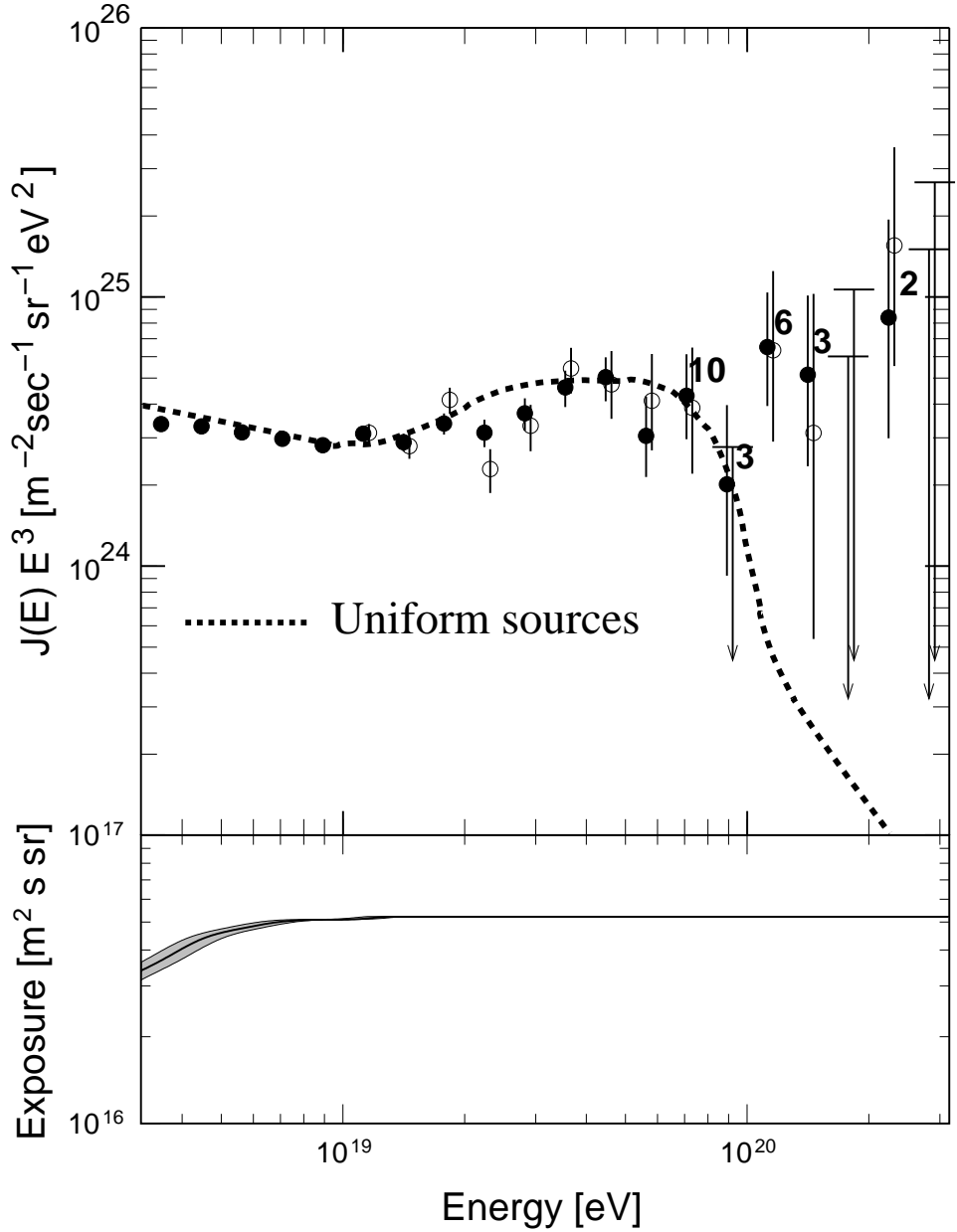}
\caption{
Energy spectrum determined by AGASA and the exposure 
with zenith angles smaller than 45$^{\circ}$ up until July 2002.
(Open circles: well contained events; Closed circles: all events)
The vertical axis is multiplied by $ E^{3} $. 
Error bars represent the Poisson upper and lower limits 
at $ 68 \% $ confidence limit and arrows are $ 90 \% $ C.L. upper limits. 
Numbers attached to the points show the number of events 
in each energy bin.
The dashed curve represents the spectrum expected for 
extragalactic sources distributed uniformly in the Universe, 
taking account of the energy determination error.
The uncertainty in the exposure is shown by the shaded region. 
}
\label{fig:spec01073_fc}
\end{figure}

Figure \ref{fig:spec01073_fc} shows the energy spectrum observed with AGASA 
with zenith angles smaller than 45$^{\circ}$ 
up until July 2002. 
The exposure (the aperture $\times$ the observation time) 
is also drawn in Figure \ref{fig:spec01073_fc} 
and is almost constant at 5.1 $\times$ 10$^{16}$m$^{2}$ s sr 
above 10$^{19}$eV 
for the events inside the array boundary. 
Closed circles indicate these ``inside'' events, and 
open circles are ``well contained'' events 
whose cores are located at least 1km inside the array boundary. 
The energy spectra for inside and well contained events 
agree well with each other 
and hence our criterion of selecting all events inside the boundary can be
justified.

Though we have examined the systematic errors 
in energy determination carefully, 
it is not easy to calibrate the absolute energy experimentally 
to decide whether 10$^{20}$eV candidate events really
exceed the GZK cutoff energy. 
One method is to compare the spectrum with the extension of Akeno energy spectrum 
measured at lower energies.
At Akeno
there are arrays of various detector-spacing 
depending on the primary energy of interest,
and energy spectra have been determined systematically 
over five decades in energy 
under the similar experimental procedures \cite{nagano84b}.

\begin{figure}
\centerline{
\includegraphics[width=0.65\textwidth]{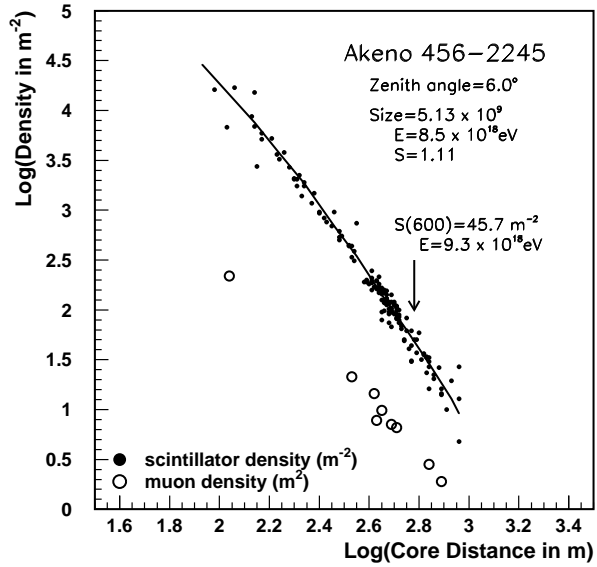}
}
\caption{
   Comparison of energies determined from $N_e$ and
   $S(600)$ for one of the largest events landing inside the
   1km$^2$ array.
}
\label{fig:compA1AGASA}
\end{figure}

\begin{figure}
\centerline{
\includegraphics[width=0.65\textwidth]{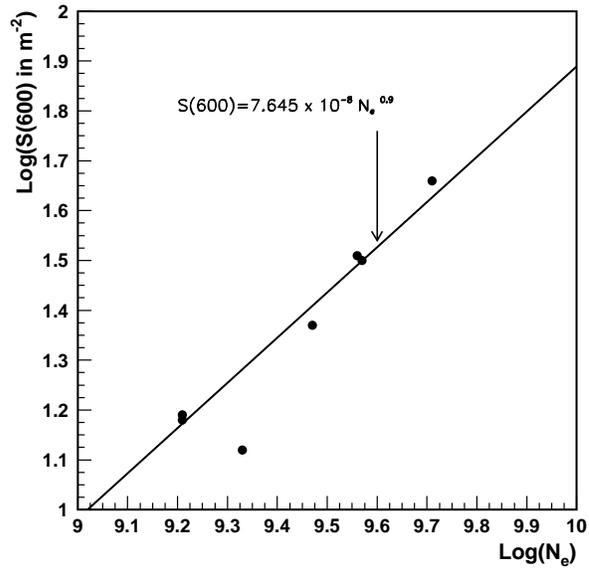}
}
\caption{
$S(600)$ and $N_{e}$ for the events landing well inside
the 1km$^2$ array. 
A solid line is $S(600)$-$N_{e}$ relation derived from 
Equations (\ref{eq:econv}) and (\ref{eq:e_ne}).
}
\label{fig:ne_s600}
\end{figure}

In the 10$^{18}$eV energy region, 
a comparison of energy determination using $S(600)$ and $N_e$ 
for each event can be made with the 1km$^2$ array, 
where 156 detectors of 1m$^2$ area
each are arranged with 120m separation.
The relation converting $N_{e}$ to energy at Akeno 
is determined exprimentally
via the longitudinal development curve measured at
 Chacaltaya and Akeno \cite{nagano84b} and is expressed by
\begin{equation}
	E_{0} \mbox{[eV]} = 3.9 \times 10^{15} 
		\left(\frac{N_{e}}{10^{6}} \right)^{0.9}
	\hspace{2em} . 
	\label{eq:e_ne}
\end{equation}
One of the largest events hitting the 1km$^2$ array
is shown in Figure \ref{fig:compA1AGASA}, 
with energies estimated from the shower size $N_{e}$ and $S(600)$.
In Figure \ref{fig:ne_s600}, 
some events are plotted on $S(600)$ vs $N_{e}$
diagram with the $S(600)$-$N_{e}$ relation from 
Equations (\ref{eq:econv}) and (\ref{eq:e_ne}). 
Though the number of events above $10^{18}$ eV
is small, the difference of
energy determined using both methods is within 10\%.
In other words, the energy conversion factor from
$S(600)$ by simulation is in good agreement with
that from $N_{e}$ by experiment.

The energy determined by the 1km$^2$ array (E$_1$)
and that by the 20km$^{2}$ array (E$_{20}$), 
whose detectors are deployed with about 1km separation and
is the prototype array of AGASA, 
have also been compared in the $10^{18}$eV energy region \cite{nagano92a}.  
The ratio E$_{20}$/E$_{1}$ is 1.10 and 
the dispersion is 45\%. 
Since the median energy of the showers is 10$^{18.1}$eV, 
the error in the $S(600)$ determination by the 20km$^{2}$ array 
is rather large and hence the wide spread is reasonable. 

\begin{figure}
\centerline{
\includegraphics[width=0.9\textwidth]{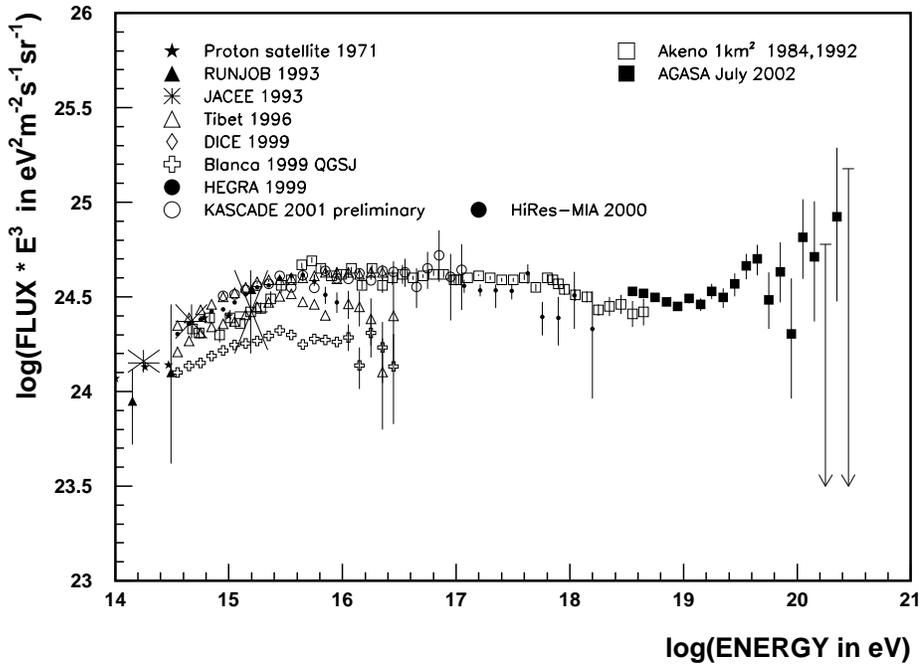}
}
\caption{
Cosmic ray energy spectrum over a  wide energy range. 
The present AGASA energy spectrum is shown by closed squares. 
The spectrum from the Akeno 1km$^2$ array is shown by open squares. 
The Akeno-AGASA energy spectrum covers more than 5 decades
of energy and is in reasonable agreement with most 
energy spectra below 10$^{18}$eV.     
}
\label{fig:spectrumAll}
\end{figure}

In Figure \ref{fig:spectrumAll}, 
the spectrum obtained by the 1km$^2$ array (E$_1$)
is shown with open squares and that by AGASA by closed squares. 
There is a difference in the overlapping energy region representing a 10\% energy difference. 
In the same figure, 
results below 10$^{18}$eV from several experiments are plotted. 
The Akeno energy spectrum is in good agreement with other experiments 
\cite{nagano00b}
from the {\it knee} to the second {\it knee} region, 
except Blanca \cite{blanca} and DICE \cite{dice}. 
The comparison of the present results with other experiments 
in the highest energy region will be made elsewhere. 

%*****************************************************************************
\section{Conclusion}

We have reevaluated the uncertainties 
in energy estimation using data accumulated over ten years. 
Table \ref{tbl:systematics} summarizes the major systematics and uncertainties 
in energy estimation. 
Here, the symbol ``$+$'' means that currently assigned energies 
should be pushed up under a particular effect, 
and the symbol ``$-$'' represents a shift in the opposite direction. 
The probable overestimation of 10\% 
due to  shower front structure and delayed particles 
may be compensated for by
the probable underestimation of the energy conversion factor by 10\%, 
an effect resulting from the inclusion of the average altitude of AGASA 
and the proper definition of what is meant by a ``single particle''.
Adding uncertainties in quadrature,
the systematic uncertainty in energy determination 
in the AGASA experiment 
is estimated to be $\pm$18\% in total. 
Therefore, the currently assigned energies of the AGASA events 
have an accuracy of 
$\pm$25\% in event-reconstruction resolution and 
$\pm$18\% in systematics. 

It should be noted that 
the Akeno-AGASA spectra cover over five decades in energy, 
connecting smoothly from the {\it knee} to a few times 10$^{20}$eV, 
except for a 10\% difference in energy in the 10$^{19}$eV region. 
This may be due to the difference in
the energy conversion relations for the experiments 
and is within the systematic errors evaluated here.
It is concluded that 
there are surely events above 10$^{20}$eV and 
the energy spectrum extends up to a few times 10$^{20}$eV. 
The present highest energy event may only be limited by exposure. 
The next generation of experiments with much larger exposures are 
highly anticipated. 

%*****************************************************************************
\section*{Acknowledgements}

We are grateful to Akeno-mura, Nirasaki-shi, Sudama-cho, Nagasaka-cho, 
Ohizumi-mura, Tokyo Electric Power Co. and Nihon Telegram and Telephone Co. 
for their kind cooperation. 
We are indebted to other members of the Akeno group 
in the maintenance of the AGASA array. 
We also thank Bruce Dawson for his kind advice in improvements 
of this article. 
This work is supported in part by JSPS 
(Japan Society for the Promotion of Science) grants in aid
of Scientific Research \#12304012.

%*****************************************************************************
%\pagebreak

%*****************************************************************************
%	FIGURES
%*****************************************************************************
\clearpage

%%% 2. Scintillation Detector %%%%%%%%%%%%%%%%%%%%%%%%%%%%%%%%%%%%%%%%%%%%%%%%

%%% 3.1 Detector %%%%%%%%%%%%%%%%%%%%%%%%%%%%%%%%%%%%%%%%%%%%%%%%%%%%%%%%%%%%%

%%% 3.2 Air Shower Phenomenology %%%%%%%%%%%%%%%%%%%%%%%%%%%%%%%%%%%%%%%%%%%%%

\begin{figure}
\centerline{
\includegraphics[width=0.65\textwidth]{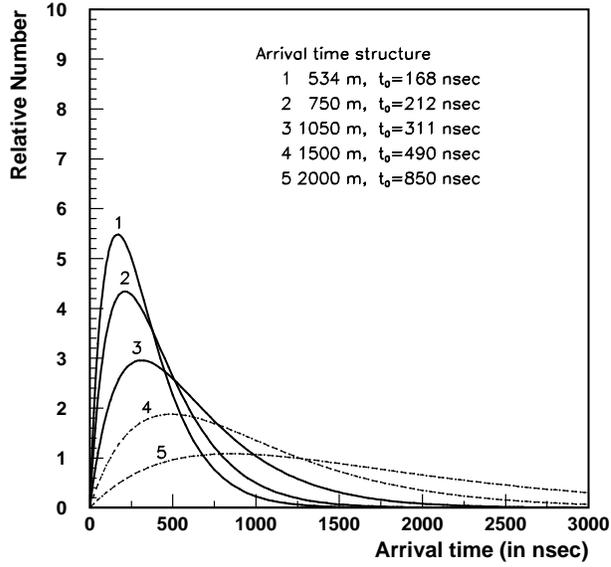}
}
\caption{
Arrival time distribution of shower particles. 
$t_{0}$ for 534m, 750m and 1,050m are determined experimentally 
with a 30m$^{2}$ scintillator \cite{khonda87a}, 
and those for 1,500m and 2,000m are extrapolated 
assuming $\log t_{0} \propto r$. 
}
\label{fig:arrtime}
\end{figure}

\begin{figure}
\centerline{
\includegraphics[width=0.65\textwidth]{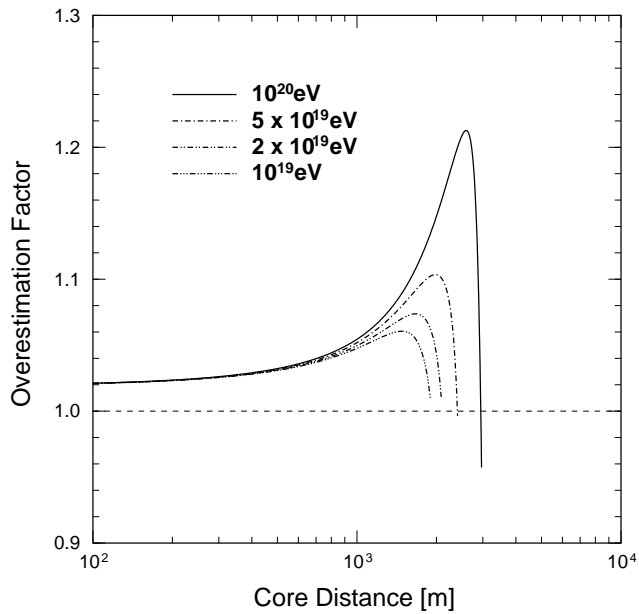}
}
\caption{
Density overestimation due to the effect of shower front thickness 
estimated by considering the LDF profile for different energies. 
The drop at large core distance occurs at a radius where the expected
particle count in a detector is one.
}
\label{fig:eff_thickness_ldf}
\end{figure}

\begin{figure}
\centerline{
\includegraphics[width=0.65\textwidth]{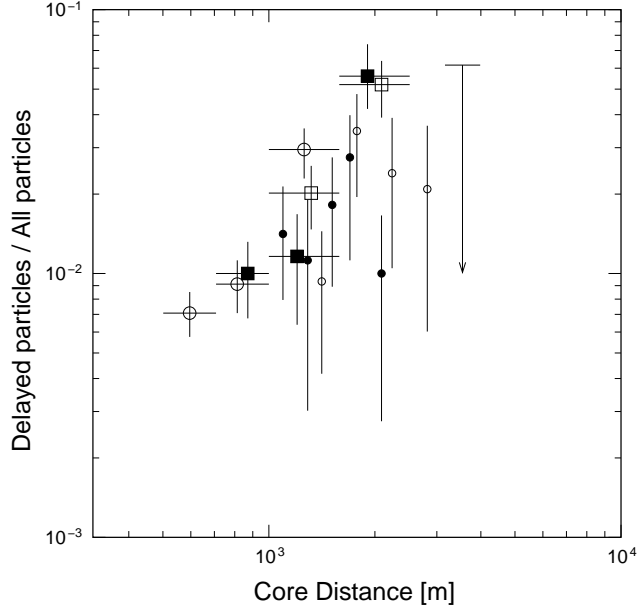}
}
\caption{
Fraction of  delayed particles to all shower particles. 
The current results are for 
$t_D \ge 3\mu$s and $N_D \ge$1.0 with three energy ranges of 
$\log$(Energy [eV]) $=$ 18.5--19.0 (big open circles), 
19.0--19.4 (big open squares) and above 19.4 (big closed squares). 
The previous Akeno result by Teshima et al. \cite{teshima86b} 
(small open circles), 
and the result by Linsley \cite{linsley} (small closed circles) are for 
$t_D \ge 4\mu$s and $N_D \ge$3.0 with energies around 10$^{18}$eV. 
}
\label{fig:NdvsE}
\end{figure}

\begin{figure}
\centerline{
\includegraphics[width=0.65\textwidth]{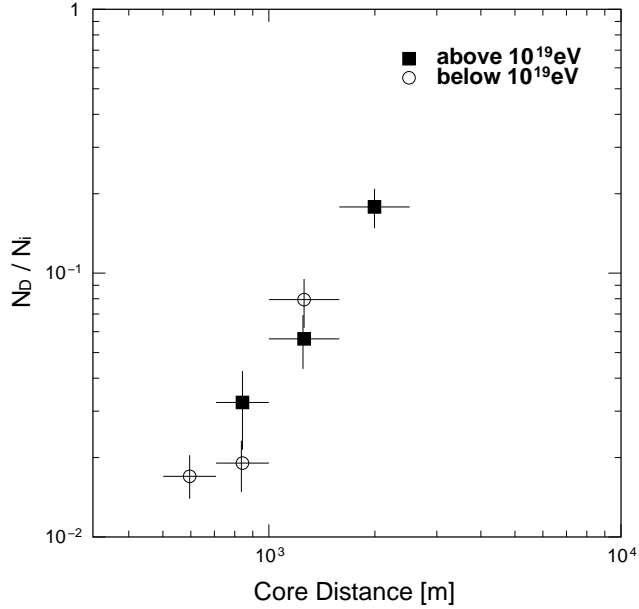}
}
\caption{
Core distance dependence of the ratio $N_{D}/N{i}$. 
}
\label{fig:delay_nd_ni}
\end{figure}

\begin{figure}
\includegraphics[width=\textwidth]{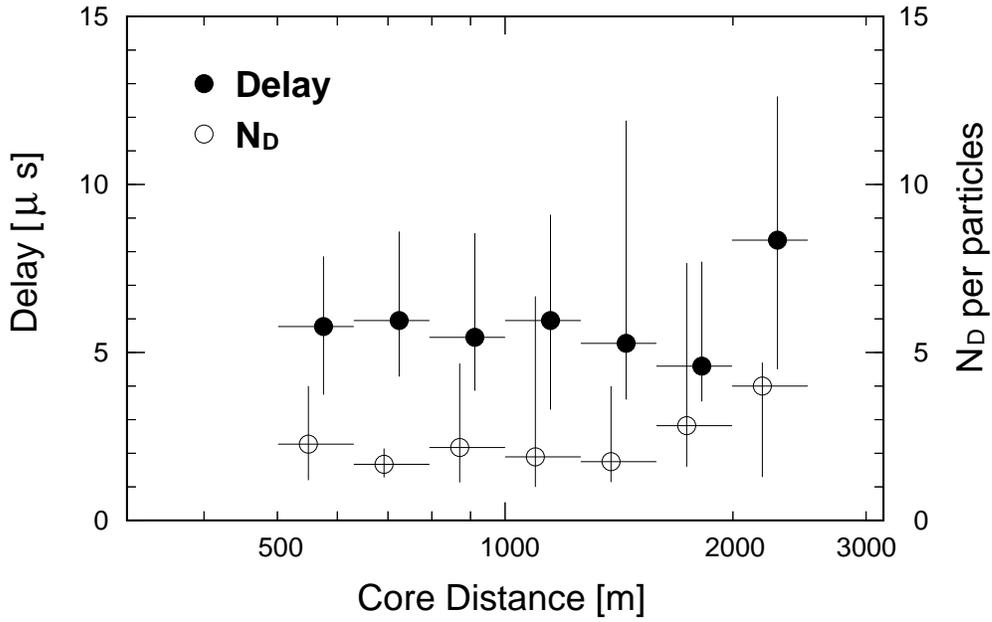}
\caption{
Delay time and $N_{D}$ of delayed particles 
as a function of core distance. 
Vertical bars indicate the 68\% confidence limits in each bin. 
Data points within each bin range from 15 to 30, 
except for largest core distance bin (3 data points). 
}
\label{fig:dis_delay_nd}
\end{figure}

\begin{figure}
\includegraphics[width=0.9311\textwidth]{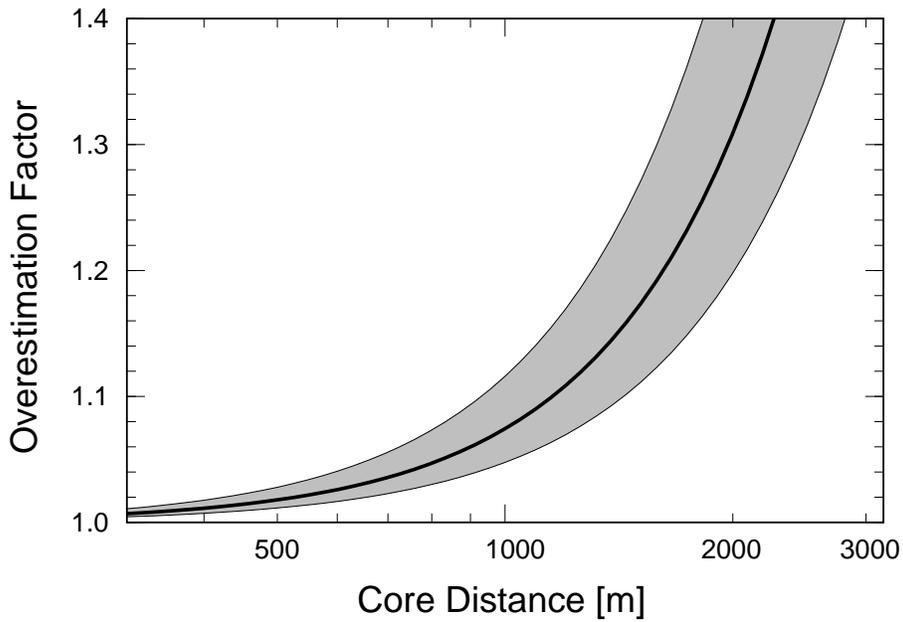}
\caption{
Overestimation factor due to delayed particles 
as a function of core distance. 
The solid line indicates the $OF$ values at each core distance 
with $t_{D} = 6\mu$s 
and their uncertainties are shown by the shaded region. 
}
\label{fig:dis_of}
\end{figure}

%%% 3.3 Energy Estimator %%%%%%%%%%%%%%%%%%%%%%%%%%%%%%%%%%%%%%%%%%%%%%%%%%%%%
%%% 3.4 Analysis %%%%%%%%%%%%%%%%%%%%%%%%%%%%%%%%%%%%%%%%%%%%%%%%%%%%%%%%%%%%%

%%% 4. Evaluation of Total Uncertainties %%%%%%%%%%%%%%%%%%%%%%%%%%%%%%%%%%%%%
%%% 5. Density Map and Lateral Distribution of 10^20eV Events %%%%%%%%%%%%%%%%
%%% 6. Energy Spectrum with \theta \leq 45^\circ %%%%%%%%%%%%%%%%%%%%%%%%%%%%%

%%% 7. Discussion %%%%%%%%%%%%%%%%%%%%%%%%%%%%%%%%%%%%%%%%%%%%%%%%%%%%%%%%%%%%
%*****************************************************************************
\setcounter{figure}{13}
\begin{figure}
\centerline{
\includegraphics[width=0.7\textwidth]{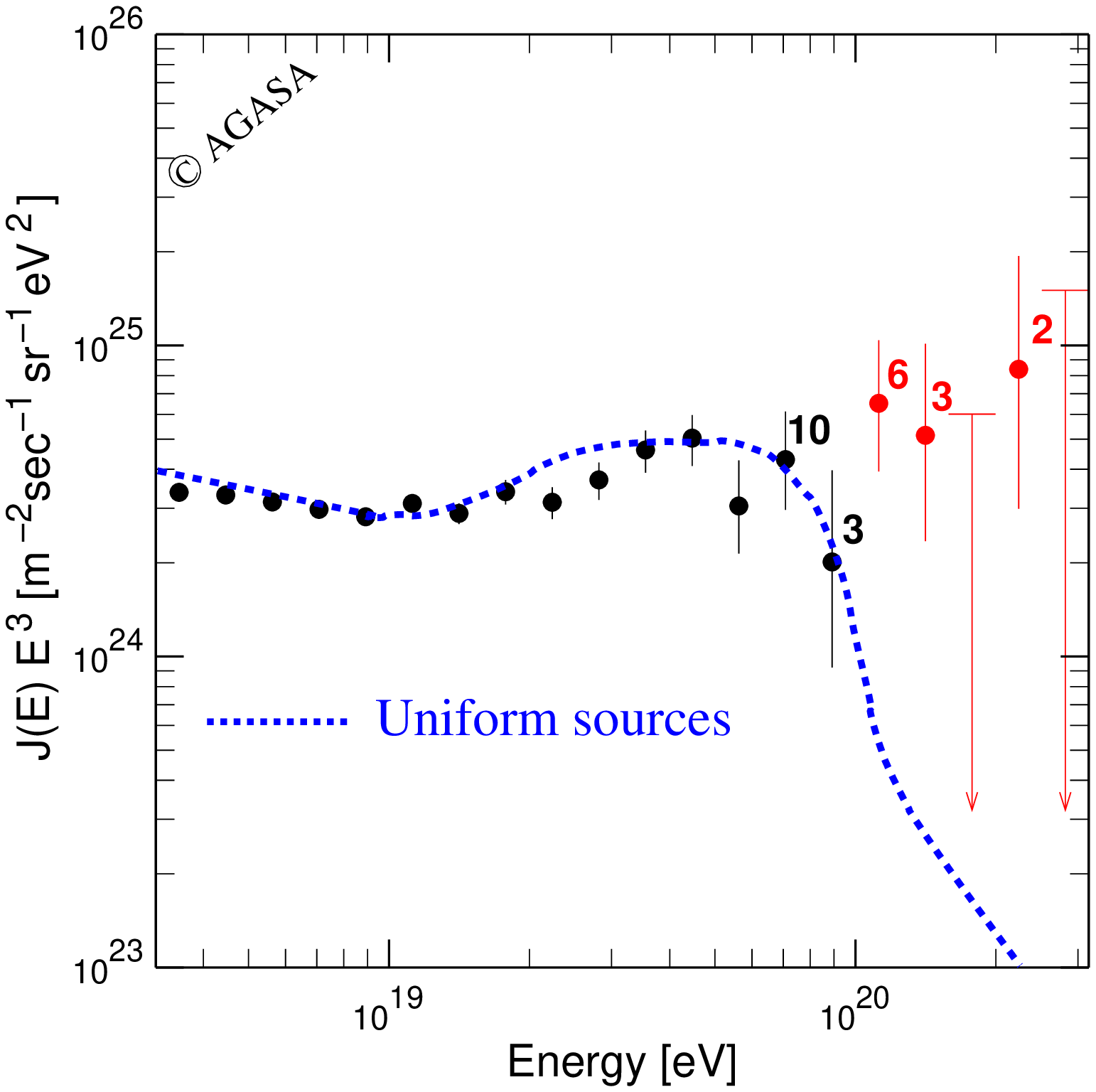}
}
%\caption{
Fig.~\ref{fig:spec01073_fc}(b).
~Same plot but only the spectrum.
%}
\end{figure}
%*****************************************************************************
%	TABLES
%*****************************************************************************
\clearpage

\begin{table}
\caption{Energy conversion from $S(600)$. 
The column  ``Single Particle'' describes
the definition of ``a single particle''
used in the evaluation of $S(600)$. 
Each formula is evaluated at the altitude given in the column ``Altitude''. 
}
\label{tbl:econv}
\begin{tabular}{lllllrcc}\hline
Simulation & Single & Altitude & Interaction & Primary 
  & \multicolumn{2}{c}{$ E = a \times$ 10$^{17} \cdot S_{0}(600)^{b}$} 
  & Citation	\\
Code       & Particle &     & Model       & Composition
  & \multicolumn{1}{c}{$a$} & \multicolumn{1}{c}{$b$} &  \\
\hline
% Dai et al.\cite{dai88a}
COSMOS  & ``electrons'' & 900m & QCDJET & p   & 2.03 & 1.02 & \cite{dai88a} \\
\hline
% Nagano et al.\cite{nagano00a}
CORSIKA & $PH_{peak}^{0}$ & 900m & QGSJET98 & p   & 2.07 & 1.03 & \cite{nagano00a}	\\
(v5.623)&                 &      &        & Fe  & 2.24 & 1.00 &	\\
        &                 &      & SIBYLL1.6 & p   & 2.30 & 1.03 &	\\
        &                 &      &        & Fe  & 2.19 & 1.01 &	\\
\hline
% Sakaki et al.\cite{sakaki01a}
AIRES   & $PW_{peak}^{\theta}$ & 667m & QGSJET98 & p   & 2.17 & 1.03 & \cite{sakaki01a}	\\
(v2.2.1)&                      &      &        & Fe  & 2.15 & 1.01 &	\\
        &                      &      & SIBYLL1.6 & p   & 2.34 & 1.04 &	\\
        &                      &      &        & Fe  & 2.24 & 1.02 &	\\
\hline
\end{tabular}
\end{table}

\begin{table}
\caption{Major systematics of AGASA.}
\label{tbl:systematics}
\begin{tabular}{p{0.05\textwidth}lr}\hline
\multicolumn{2}{l}{Detector:}				&		\\
& detector absolute gain				& $\pm$ 0.7\%	\\
& detector linearity					& $\pm$ 7\%	\\
& detector response (box, housing, ...)			& $\pm$ 5\%	\\
\multicolumn{2}{l}{Air shower phenomenology:}		&		\\
& lateral distribution function				& $\pm$ 7\%	\\
& $S(600)$ attenuation					& $\pm$ 5\%	\\
& shower front structure		& $-$ 5\% $\pm$ 5\%	\\
& delayed particles			& $-$ 5\% $\pm$ 5\%	\\
\multicolumn{2}{l}{Energy estimator $S(600)$:}	 	&		\\
& $	\left.	\begin{array}{l}
	\mbox{interaction models, chemical compositions (p/Fe),}	\\
	\mbox{simulation codes, height correction,}			\\
	\mbox{$S(600)$ fluctuation}
		\end{array}
	\right\} $				& $+$10\% $\pm$12\%	\\
\hline
\multicolumn{2}{l}{Total}				& $\pm$18\%	\\
\hline
\end{tabular}
\end{table}

%*****************************************************************************
\end{document}